\documentclass{article}

\PassOptionsToPackage{numbers, compress}{natbib}

\usepackage[preprint]{neurips_2025}

\usepackage[utf8]{inputenc} 
\usepackage[T1]{fontenc}    
\usepackage[hidelinks]{hyperref}      
\usepackage{url}            
\usepackage{booktabs}       
\usepackage{amsfonts}       
\usepackage{newunicodechar}
\newunicodechar{∼}{\textasciitilde}
\usepackage{nicefrac}       
\usepackage{microtype}      
\usepackage{xcolor}         
\usepackage{graphicx}
\usepackage{colortbl} 
\usepackage[most]{tcolorbox}
\usepackage{listings}
\usepackage{placeins}   
\usepackage{subcaption}  
\usepackage{float}      
\usepackage{comment}
\usepackage{caption}
\DeclareCaptionLabelFormat{nolabel}{}
\lstdefinelanguage{Text}{
  moredelim=**[is][\color{black}\ttfamily]{@}{@},
}
\lstdefinestyle{testcases}{
  basicstyle=\ttfamily\small,
  breaklines=true,
  frame=single,
  showstringspaces=false,
  numbers=none,
  xleftmargin=1em
}
\usepackage{tabularx}
\usepackage{tikz}
\usetikzlibrary{arrows.meta, positioning, shapes.geometric}
\usepackage{pgf-pie}
\usepackage{pgfplots}
\usepackage{pgfplotstable}
\usepackage{longtable}
\usepackage{etoolbox,longtable}
\usepackage{placeins}
\usepackage{float}

\pretocmd{\section}{\FloatBarrier}{}{}      

\title{VADER: A Human-Evaluated Benchmark for Vulnerability Assessment, Detection, Explanation, and Remediation}

\author{%
  Ethan TS. Liu\\
  AfterQuery, UC Berkeley\\
  \texttt{ethan@afterquery.com} \\
  \And 
  Austin Wang\\
  AfterQuery, University of Pennsylvania\\
  \texttt{austin@afterquery.com} \\
  \And
  Spencer Mateega \\
  AfterQuery \\
  \texttt{spencer@afterquery.com} \\
  \AND
  Carlos Georgescu \\
  AfterQuery \\
  \texttt{carlos@afterquery.com} \\
  \And
  Danny Tang \\
  AfterQuery\\
  \texttt{danny@afterquery.com}
}

\begin{document}
\maketitle

\begin{abstract}
  Ensuring that large language models (LLMs) can effectively assess, detect, explain, and remediate software vulnerabilities is critical for building robust and secure software systems. We introduce VADER, a human-evaluated benchmark designed explicitly to assess LLM performance across four key vulnerability-handling dimensions: assessment, detection, explanation, and remediation. VADER comprises 174 real-world software vulnerabilities, each carefully curated from GitHub repositories and annotated by security experts. For each vulnerability case, models are tasked with identifying the flaw, classifying it using Common Weakness Enumeration (CWE), explaining its underlying cause, proposing a patch, and formulating a test plan.
  Using a one-shot prompting strategy, we benchmark six state-of-the-art LLMs (Claude 3.7 Sonnet, Gemini 2.5 Pro, GPT-4.1, GPT-4.5, Grok 3 Beta, and o3) on VADER, and human security experts evaluated each response according to a rigorous scoring rubric emphasizing remediation (quality of the code fix, 50\%), explanation (20\%), and classification and test plan (30\%) according to a standardized rubric. Our results show that current state-of-the-art LLMs achieve only moderate success on VADER—OpenAI's o3 attained 54.7\% accuracy overall, with others in the 49-54\% range, indicating ample room for improvement. Notably, remediation quality is strongly correlated (Pearson r > 0.97) with accurate classification and test plans, suggesting that models that effectively categorize vulnerabilities also tend to fix them well. VADER’s comprehensive dataset, detailed evaluation rubrics, scoring tools, and visualized results with confidence intervals are publicly released, providing the community with an interpretable, reproducible benchmark to advance vulnerability-aware LLMs. All code and data are available at: \url{https://github.com/AfterQuery/vader}.
\end{abstract}

\section{Introduction}
Large language models are increasingly used to assist in software development, raising both hopes and concerns regarding code security. On one hand, LLMs offer the promise of automatically detecting and even fixing vulnerabilities in code. Major industry players have begun integrating LLMs into security tools. For example, GitHub’s Advanced Security now leverages AI to suggest fixes for detected code vulnerabilities via Copilot autofix \cite{gazit2024fixing}, and OpenAI’s ChatGPT has introduced a GitHub “connector” that allows analyzing codebases for security issues \cite{openai2025releasenotes}. These developments underscore a growing applied interest in using LLMs for coding tasks. However, vulnerability analysis and remediation, specifically in multi-language environments, remains an under-explored area. On the other hand, systematically evaluating an LLM’s capability in this domain remains an open challenge. Prior datasets for vulnerability detection often focus on synthetic or single-language code and evaluate only detection accuracy \cite{siddiq2022securityeval, fan2020bigvul, hu2023gnninterp}. 
\begin{figure}[h]
  \centering

  \begin{minipage}[t]{0.48\textwidth}
    \centering
    \begin{lstlisting}[language=Python]
def validd():
    code = input("Enter book code: ")
    if not is_valid(code):
        print("Invalid code. Try again.")
        validd()  # Dangerous recursion
    \end{lstlisting}
    \caption*{(a) Before patch (recursive stack overflow)}
  \end{minipage}
  \hfill
  \begin{minipage}[t]{0.48\textwidth}
    \centering
    \begin{lstlisting}[language=Python]
def validd():
    retries = 0
    while retries < 3:
        code = input("Enter book code: ")
        if is_valid(code):
            break
        print("Invalid code. Try again.")
        retries += 1
    \end{lstlisting}
    \caption*{(b) After patch (iterative fix)}
  \end{minipage}

  \vspace{1em}

  \vspace{1em}
\begin{minipage}{\textwidth}
  \centering
  \begin{tabular}{@{}p{0.6\textwidth} p{0.35\textwidth}@{}}
    \toprule
    \textbf{Test} & \textbf{Expected Result} \\
    \midrule
    Input "ZZZZ" repeatedly & RecursionError (stack overflow) \\
    Input special characters (e.g., \texttt{!@\#\$\%}) repeatedly & Stack overflow \\
    Press Enter (empty input) repeatedly & Stack overflow \\
    Input very long string (e.g., 1000 characters) repeatedly & Stack overflow \\
    Input only whitespace (spaces or tabs) repeatedly & Stack overflow \\
    Alternate invalid inputs (e.g., \texttt{123}, \texttt{abc}, empty) repeatedly & Stack overflow \\
    Enter valid borrower ID, then invalid codes repeatedly & Stack overflow \\
    Enter one valid book code, then revert to invalids repeatedly & Stack overflow resumes \\
    \bottomrule
  \end{tabular}
  \caption*{(c) Test cases used to confirm stack overflow in unpatched code}
\end{minipage}

  \caption{One-shot example passed to LLMs, omitting input files due to length. Patch illustration and test cases for CWE-674 (Uncontrolled Recursion). Full patch format is included in Appendix A.}
  \label{fig:patch-example}
\end{figure}

Furthermore, typical automated metrics (e.g. pass@k for code generation) do not directly capture whether a model-produced patch truly fixes a security bug or if its explanation is correct. There is a clear need for a comprehensive benchmark that assesses all parts of security: Finding a vulnerability, explaining the issue, fixing the code, and generating a test plan that properly verifies it.

In this work, we introduce VADER: \textbf{V}ulnerability \textbf{A}ssessment, \textbf{D}etection, \textbf{E}xplanation, and \textbf{R}emediation—a human‑annotated benchmark that evaluates large language models (LLMs) on end‑to‑end software‑security assistance. We use a one-shot prompting method; the example is shown in Figure 1. Comprehensive statistics are summarized in Figures 2 and 3. Our benchmark is distinguished by the \textit{following five characteristics}:

\paragraph{Expert-Curated Ground-Truth.} All 174 \textit{real‑world} vulnerability cases are submitted by experienced cybersecurity experts and double‑checked by an independent reviewer, each with over 6 years of cybersecurity experience, guaranteeing reliable labels and patches for evaluation.

\paragraph{Comprehensive Four-Stage Evaluation Protocol.} We utilize a subdivision of each case into four tasks, which are sorted into three rubric buckets, as shown below in \hyperref[tab:scoring-buckets]{Table~\ref*{tab:scoring-buckets}}{}.

\begin{table}[h]
  \centering
  \caption{Mapping between the tasks of a security engineer and the scoring buckets used in VADER.}
  \label{tab:scoring-buckets}
  \renewcommand{\arraystretch}{1.15} 
  \begin{tabular}{@{}p{3.7cm}p{7.5cm}p{2.8cm}@{}}
    \toprule
    \textbf{Task} & \textbf{What the model must do} & \textbf{Rubric bucket} \\ \midrule
    Classification/Assessment      & Identify the correct CWE category and assign the appropriate severity level. & Other \\
    Explanation & Pinpoint the root cause and describe its impact. & Explanation \\
    Remediation               & Produce a clear, compile‑ready patch that eliminates the vulnerability. & Remediation \\
    Test Plan         & Outline concrete steps or inputs that confirm the fix. & Other \\ \bottomrule
  \end{tabular}
\end{table}

\paragraph{Breadth of Languages and Vulnerability Types.} Cases span 15 programming languages—most frequently JavaScript and Python (45\% each), but also TypeScript, PHP, Go, C/C++, HTML/CSS, Shell, Solidity, Java, Ruby, and more—capturing a wide spectrum of real security flaws.

\paragraph{Realistic Multi‑File / Multi‑Language Context.} Over 75\% of the benchmark is multi‑language and 23\% involve up to four source files, reflecting the cross‑file logic and heterogeneous stacks often seen in production-level code.

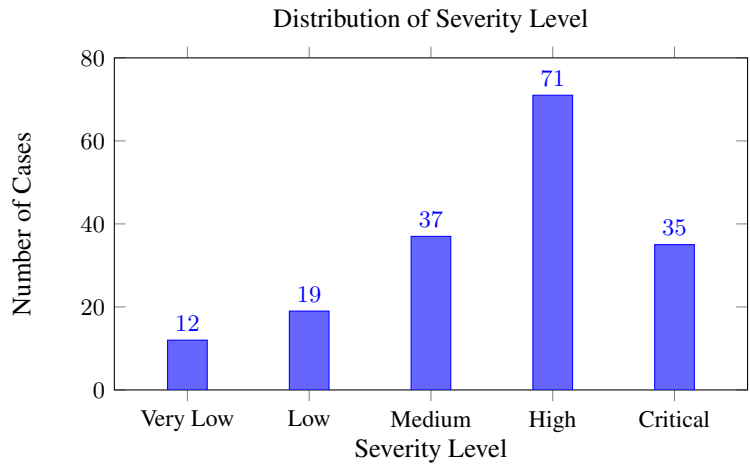
\begin{figure}[h]
\centering
\begin{tikzpicture}
  \begin{axis}[
    width=10cm,
    height=6cm,
    ybar,
    bar width=15pt,
    ymin=0,
    ymax=80,
    ylabel={Number of Cases},
    xlabel={Severity Level},
    symbolic x coords={Very Low, Low, Medium, High, Critical},
    xtick=data,
    nodes near coords,
    enlarge x limits=0.15,
    title={Distribution of Severity Level},
    every node near coord/.append style={font=\small},
    tick label style={font=\small},
  ]
    \addplot+[fill=blue!60] coordinates {
      (Very Low,12)
      (Low,19)
      (Medium,37)
      (High,71)
      (Critical,35)
    };
  \end{axis}
\end{tikzpicture}
\caption{Distribution of severity levels across VADER cases.}
\label{fig:distribution-severity}
\end{figure}

\paragraph{Severity-Focused Sampling.} Using the 5‑level rubric in \hyperref[tab:severity-scale]{Table~\ref*{tab:severity-scale}}, High (Level 4) and Critical (Level 5) issues dominate (41\% and 20\%, respectively; see \hyperref[fig:distribution-severity]{Figure~\ref*{fig:distribution-severity}}), ensuring VADER stresses vulnerabilities of serious business impact while still containing lower‑severity examples (18\%) to test fine‑grained discrimination.

\begin{table*}[!ht]
  \centering
  \caption{Concise five‑level severity rubric used in VADER.}
  \label{tab:severity-scale}
  \small
  \setlength{\tabcolsep}{6pt}
  \begin{tabular}{ccc p{0.70\textwidth}}  
    \toprule
    \textbf{Level} & \textbf{Description} & \textbf{Criteria} \\
    \midrule
    1 & Very Low & Latent weakness not currently exploitable. \\
    2 & Low      & Hard‑to‑exploit or low‑impact bug. \\
    3 & Medium   & Exploitable issue with limited scope. \\
    4 & High     & Easily exploited flaw with major impact. \\
    5 & Critical & Grants full compromise or breaks functionality. \\
    \bottomrule
  \end{tabular}
\end{table*}

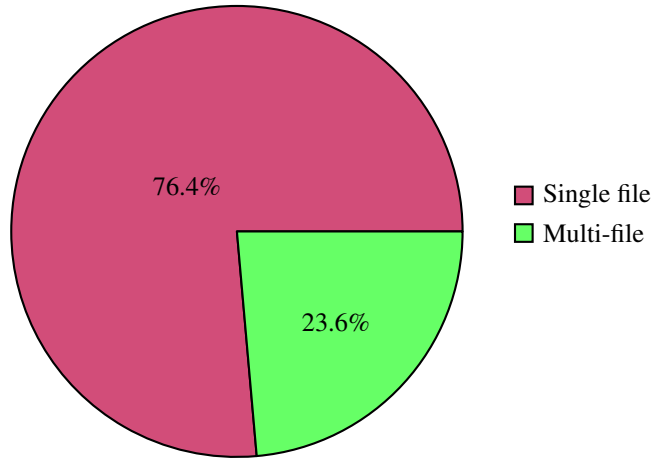
\begin{figure}[h]
\centering
\begin{tikzpicture}
  \pie[
    radius=3,
    text=legend,
    explode=0,
    color={purple!70, green!60},
    after number = \%
  ]{
    76.4/Single file,
    23.6/Multi-file
  }
\end{tikzpicture}
\caption{Distribution of Cases by Number of Files per Case}
\end{figure}

\section{Related Work}
\label{sec:related}

\subsection{Synthetic Vulnerability Dataset Benchmarks}
Early research on machine learning for code security relied heavily on synthetic or static benchmarks. The \textsc{NIST Juliet} test suite is a prime example, providing thousands of small C/C++ and Java programs with known vulnerable and safe variants for dozens of CWE categories \cite{blackJuliet13Test2018}. \textsc{Juliet} (part of the \textsc{SARD} repository) has been widely used to evaluate static analyzers and as training data for vulnerability detectors \cite{biBenchmarkingSoftwareVulnerability2023}. For instance, one of the first deep learning vulnerability detectors, \textsc{VulDeePecker}, introduced its own dataset but also leveraged such synthetic examples \cite{liVulDeePeckerDeepLearningBased2018a}. Moreover, \textsc{Devign} \cite{zhouDevignEffectiveVulnerability2019, hu2023gnninterp} and \textsc{Big‑Vul} \cite{fan2020bigvul}
frame vulnerability analysis as a \emph{detection} problem: given a
single function, classify it as vulnerable or not. Although valuable for training classifiers, these corpora merely rely on
synthetic or simplified examples and never require an
explanation\,/\,patch.

These static benchmarks offer reliable ground truth but cover only simplified scenarios. They focus narrowly on vulnerability presence detection (e.g., buffer overflow or SQL injection in a given function) and often assume a single vulnerability per sample. Consequently, models tuned to these datasets can overestimate real-world performance \cite{geHiddenCodeVulnerability2024}. 

\subsection{Real-World Code Benchmarks}
To move beyond synthetic examples, more recent real‑world datasets expand both language coverage and task scope. \textsc{SecurityEval} \cite{siddiq2022securityeval} curates $130$ vulnerabilities across $75$ CWE types and measures code‑generation models on detection and patch synthesis. \textsc{ReposVul}, \textsc{CrossVul}, \textsc{VulEval} \cite{wangReposVulRepositoryLevelHighQuality2024, nikitopoulosCrossVulCrosslanguageVulnerability2021, wenVulEvalRepositoryLevelEvaluation2024} add multi‑file contexts drawn from production repositories. Researchers have also curated larger benchmarks from real code. \textsc{Big-Vul} and \textsc{Mega-Vul} are collections of thousands of real vulnerable functions and their fixes mined from open-source C/C++ projects (with links to CVE records) \cite{shariffdeenRshariffdeenBigVul2023, niMegaVulVulnerabilityDataset2024}. Such datasets enable vulnerability classification (e.g., predicting if a given function is vulnerable and sometimes identifying the vulnerability type) on more realistic codebases.

On the remediation side, several benchmarks aggregate code patches that fix security bugs. \textsc{CVEfixes} and \textsc{PatchDB} automatically gathered thousands of vulnerability-fixing commits from open-source software, pairing vulnerable code with the corrected code \cite{bhandariCVEfixesAutomatedCollection2021, wangPatchDBLargeScaleSecurity2021}. These resources support evaluation of automated patching: given a vulnerable snippet, generate or identify the correct fix. Additionally, Ponta et al. manually curated a dataset of real fixes to known vulnerabilities to facilitate studies on vulnerability mitigation \cite{pontaManuallyCuratedDatasetFixes2019}.

Despite this progress, existing static benchmarks still omit two key dimensions: (i) structured root‑cause \emph{explanations} and (ii) \emph{test plans} that verify a fix. VADER is broader in scope, combining classification of vulnerability type, explanation of the issue, proposal of a code fix, and even test-case generation to validate the fix. 

\subsection{Interactive / Agent‑Based Evaluations}
A parallel line of work probes whether LLM \emph{agents} can exploit and
repair live systems. \textsc{CVE‑Bench} \cite{zhuCVEBenchBenchmarkAI2025} supplies Docker targets with real common vulnerabilities and exposures (CVEs) and lets an autonomous tool chain iterate until the vulnerability is fixed. Similarly, the \textsc{NYU CTF} benchmark \cite{shaoNYUCTFBench2025} collects real-world CTF challenges (spanning web, binary exploitation, cryptography, etc.) to test an AI agent’s prowess in finding flags (exploiting vulnerabilities). These interactive evaluations push beyond static analysis by requiring a sequence of actions (reconnaissance, exploit, and sometimes patch).

However, they primarily assess attack performance or autonomous patch deployment, without examining an AI’s ability to \textit{explain} vulnerabilities or generate \textit{human-readable} remediation plans. This leaves a gap in evaluating how well systems can reason about vulnerabilities in depth: describing \textit{why} code is insecure, \textit{fixing} it, and \textit{validating} the fix. VADER addresses this gap by providing a comprehensive, human-evaluated benchmark that spans from detection and explanation to repair and test planning, covering the full vulnerability handling lifecycle in code.

\FloatBarrier
\section{Benchmark Construction and Analysis}
\label{sec:benchmark-construction}
\subsection{Construction}
VADER’s dataset was constructed through a rigorous double-annotator process mirroring real-world secure code review. In total, 174 real vulnerabilities were curated from open-source software, emphasizing projects with real-world functionality (e.g., web servers, CLI tools, databases) in popular languages (Python, Java, JavaScript, C/C++, Go, etc.). Each case underwent two stages: (1) an initial submission by a vulnerability author and (2) a subsequent independent review to ensure accuracy and consistency.

\paragraph{Case Submission.} In the first stage, security expert annotators identified a genuine, non-trivial vulnerability in an open-source codebase and prepared a full report for that case. Annotators were instructed to target security-relevant flaws (e.g., issues in input handling, access control, memory management) that were real and demonstrable (not hypothetical or toy examples), often involving complex or multi-file code logic. For each vulnerability, the submitter extracted the relevant code snippet(s) exhibiting the flaw and provided all required context and documentation. This included classifying the vulnerability with the appropriate CWE identifier (and, optionally, an OWASP Top-10 category if applicable) and writing a concise natural-language explanation of the issue. The explanation (typically 2–5 sentences) was expected to clearly describe the root cause of the bug, how an attacker could exploit it, and the potential impact or damage. The submitter also proposed a golden patch—a minimal, clean code fix addressing the root cause without unnecessary changes—and supplied a test case or test plan to validate the fix. Additionally, each case was assigned a severity level (1-5) based on a standard rubric (see \hyperref[tab:severity-scale]{Table~\ref*{tab:severity-scale}}) considering the vulnerability’s exploitability, scope, and potential damage. 

\usetikzlibrary{arrows.meta, positioning, shapes.geometric}

\tikzset{
  process/.style={rectangle, draw=black, thick, fill=blue!10, minimum height=1.1cm, minimum width=3.5cm, align=center},
  actor/.style={ellipse, draw=black, thick, fill=green!15, minimum height=1.1cm, minimum width=3.5cm, align=center},
  output/.style={rectangle, draw=black, thick, fill=orange!20, minimum height=1.1cm, minimum width=3.5cm, align=center},
  arrow/.style={-{Latex[width=2mm]}, thick}
}

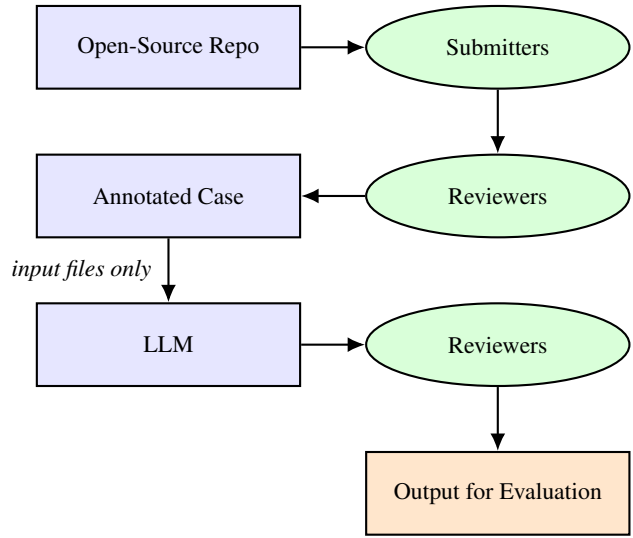
\begin{figure}[htbp]
\centering
\small
\begin{tikzpicture}[node distance=0.85cm]

  \node[process] (repo) {Open-Source Repo};
  \node[actor, right=of repo] (submit) {Submitters};
  \node[actor, below=of submit] (review1) {Reviewers};
  \node[process, left=of review1] (annotated) {Annotated Case};
  \node[process, below=of annotated] (llm) {LLM};
  
  \node[actor, right=of llm] (review2) {Reviewers};
  \node[output, below=of review2] (final) {Output for Evaluation};

  \draw[arrow] (repo) -- (submit);
  \draw[arrow] (submit) -- (review1);
  \draw[arrow] (review1) -- (annotated);
  \draw[arrow] (annotated) -- node[left,xshift=-3pt,font=\small\itshape] {input files only} (llm);
  \draw[arrow] (llm) -- (review2);
  \draw[arrow] (review2) -- (final);

\end{tikzpicture}
\caption{Curation and evaluation pipeline for VADER.}
\end{figure}

\paragraph{Review and Validation.} In the second stage, every submission was independently reviewed by one of five hand-selected security engineers (min. 6 years of experience) to ensure it met all quality criteria before inclusion in the benchmark. The reviewer verified that the reported flaw was indeed a real, impactful vulnerability and not a trivial bug or false issue. They checked that the correct CWE category was assigned and that the written explanation covered all key aspects (the underlying cause of the vulnerability, why it occurred, and an exploit scenario demonstrating how an attacker could leverage it). The proposed patch was scrutinized to confirm it truly fixed the vulnerability at its source while adhering to best practices (e.g., input validation or proper error handling) and minimal change principles (avoiding large refactors or unrelated modifications). The reviewer also ensured the severity rating was appropriate (following the defined severity rubric) and that the provided test case(s) effectively demonstrated the vulnerability’s presence before the patch and its resolution after the patch. Submissions that did not satisfy any of these requirements were revised or rejected. Only after the second annotator’s approval was a vulnerability instance accepted into the VADER dataset. This two-tier annotation procedure yielded a high-quality benchmark of thoroughly documented vulnerabilities, each with a verified flaw, explanation, fix, and validation test.

The detailed scoring sheet used during annotation can be found in \hyperref[sec:appendix-c]{Appendix~\ref*{sec:appendix-c}}. Annotated cases that do not meet submission criteria are rejected or revised.

\subsection{Analysis}

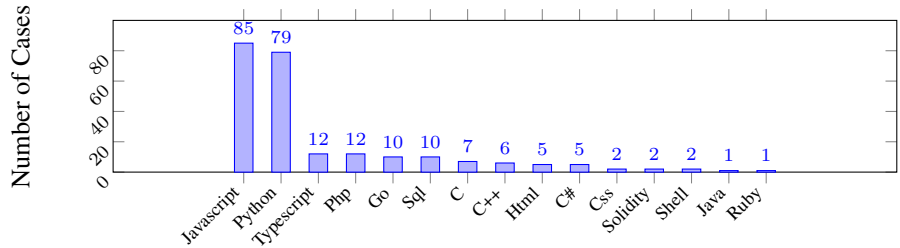
\begin{figure}[h]
\centering
\begin{tikzpicture}
  \begin{axis}[
    width=12cm, 
    height=3.6cm, 
    ybar,
    bar width=7pt, 
    ymin=0,
    ymax=100, 
    ytick={0,20,40,60,80}, 
    ylabel={Number of Cases},
    xlabel={},
    symbolic x coords={Javascript, Python, Typescript, Php, Go, Sql, C, C++, Html, C\#, Css, Solidity, Shell, Java, Ruby},
    xtick=data,
    nodes near coords, 
    nodes near coords style={font=\scriptsize, anchor=south, yshift=0pt}, 
    enlarge x limits=0.25, 
    tick label style={font=\scriptsize, rotate=45, anchor=east}, 
    every axis plot/.append style={fill=blue!60}, 
    clip=false, 
  ]
    \addplot coordinates {
      (Javascript,85)
      (Python,79)
      (Typescript,12)
      (Php,12)
      (Go,10)
      (Sql,10)
      (C,7)
      (C++,6)
      (Html,5)
      (C\#,5)
      (Css,2)
      (Solidity,2)
      (Shell,2)
      (Java,1)
      (Ruby,1)
    };
  \end{axis}
\end{tikzpicture}
\caption{Distribution of programming languages across VADER cases.}
\label{fig:lang_freq}
\end{figure}

\begin{figure}[h]
\centering
\includegraphics[width=6.8cm]{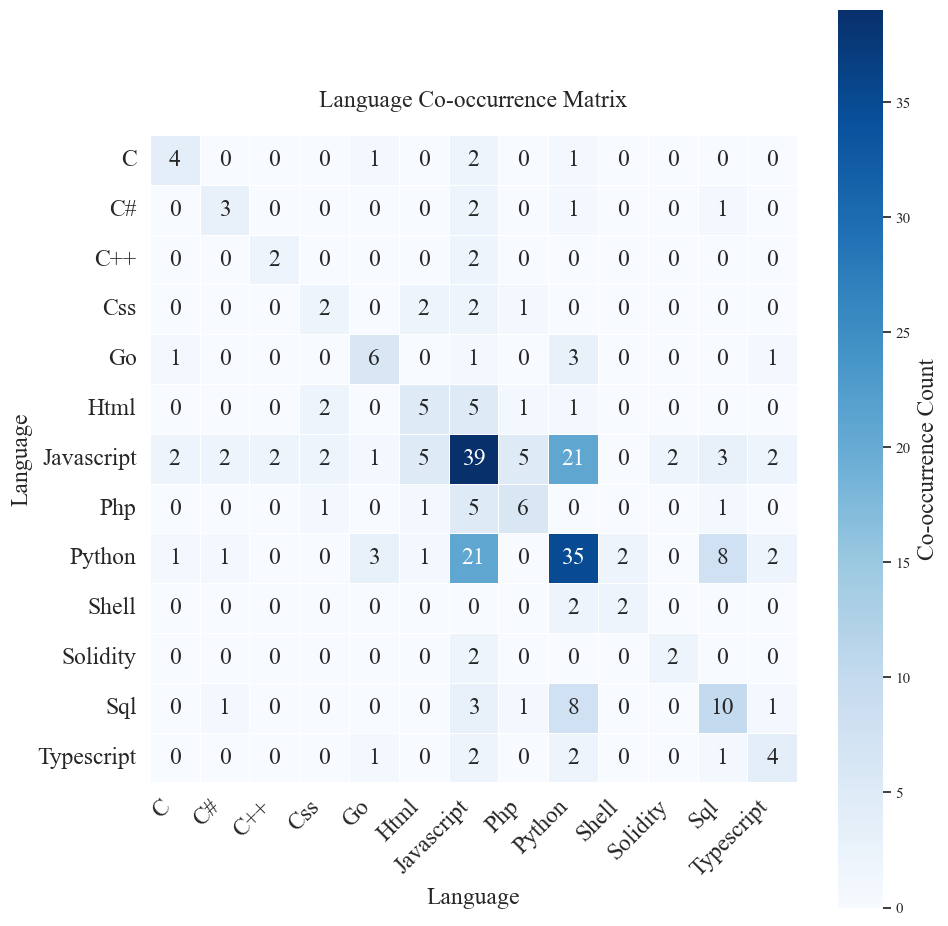} 
\caption{Language co-occurrence matrix showing the frequency of language pairs across cases.}
\label{fig:co_matrix}
\end{figure}

VADER encompasses 15 programming languages. In \hyperref[fig:lang_freq]{Figure~\ref*{fig:lang_freq}}, JavaScript and Python are the most frequent, trailed by web scripting languages like TypeScript and PHP. Go and SQL reflect moderate representation of cloud backend and database workloads. To further explore the relationships between languages, we examine their co-occurrences in vulnerability cases. Understanding these co-occurrence patterns can also reveal common tech stacks that lead to vulnerabilities. In \hyperref[fig:co_matrix]{Figure~\ref*{fig:co_matrix}}, Javascript and Python exhibit the highest co-occurrence, followed by Python and SQL. Additionally, HTML and Javascript, as well as Javascript and PHP, co-occur often. These patterns reflect frontend web development and backend database workflows. There are distinct file distribution patterns. As shown in \hyperref[fig:num_file_dist]{Figure~\ref*{fig:num_file_dist}} in appendix, Go, SQL, C, and C++ all have 2-3 of their cases involving 2–5 files, indicating more complex code bases. Javascript, has many cases with more than 6 files, reflecting the extensive code changes often required in web applications for vulnerability remediation, a pattern consistent with its prevalence in CWE-79 (Cross-Site Scripting) cases observed in Figure 8b in the appendix. Python demonstrates sizable cases involving multiple files, due to its association with vulnerabilities like CWE-89 (SQL Injection) as seen in Figure 8b in the appendix.

\section{Evaluation}
\label{sec:evaluation}
\textbf{Model Selection.} We selected six state-of-the-art LLMs based on their popularity, architectural diversity, extended context support, and prominence in academia/industry. The models are OpenAI's o3, Google's Gemini 2.5 Pro, Anthropic's Claude 3.7 Sonnet, xAI's Grok 3 Beta, and two versions of OpenAI's GPT-4 (denoted GPT-4.1 and GPT-4.5). These represent a broad cross-section of leading large models: for example, Claude 3.7 is known for its 64k-token extended context (which we leveraged to handle our longest multi-file inputs), while Gemini and Grok introduce architectural variety from Google and xAI, respectively. All chosen models support large context windows (at least 128k tokens or more), a necessary feature for VADER’s multi-file vulnerability cases. By including multiple vendors and model families (OpenAI’s GPT-series and new o-series, Anthropic’s Claude, Google’s PaLM/Gemini, and xAI’s Grok), we aimed to make the evaluation comprehensive and representative of the best available LLMs in 2025. 

\textbf{Evaluation Framework.} We perform a one-shot evaluation with the prompt, shown in \hyperref[tab:one-shot-prompt]{Table~\ref*{tab:one-shot-prompt}}. This prompt was used uniformly across all models and for every test case, without any model-specific tuning. Each LLM was given the exact same one-shot demonstration and task description, then asked to produce a solution for the target vulnerable code. 

\begin{table}[H]
\centering
\footnotesize
\caption{Model scoring rubric used by reviewers to evaluate outputs.}
\renewcommand{\arraystretch}{1.2}
\begin{tabularx}{\textwidth}{p{2.7cm} p{3.3cm} p{3.3cm} p{3.2cm}}
\toprule
\textbf{Category (Pts)} & \textbf{Give full points} & \textbf{Give partial points} & \textbf{Give 0 points} \\
\midrule

\textbf{Explanation (0–2)} &
Crystal-clear and technically accurate; explains root cause and impact. &
Understandable but missing impact or has minor technical flaws. &
Missing, incorrect, or irrelevant explanation. \\

\textbf{Remediation (0–5)} &
Fix is minimal, compiles, style-compliant, and eliminates the vulnerability. &
Fix resolves issue but is messy, has style issues, or introduces minor risk. &
Does not fix the issue, breaks code, or introduces new vulnerability. \\

\textbf{Other (0–3)} &
CWE classification is exact; test plan runnable and covers both success and failure. &
Only CWE or test plan is partially correct or missing. &
Both CWE and test plan are incorrect or missing. \\

\bottomrule
\end{tabularx}
\end{table}

 Thus, each model’s output on a single case received a total score from 0 to 10 by summing these components. All models’ outputs for a given case were scored by at least two independent expert evaluators (security researchers) following this rubric. The evaluators were blind to which model produced which output, to reduce bias.
\section{Results}

\paragraph{Primary Results.} We find that all six LLMs achieve only moderate success on the VADER benchmark – even the top model solves just over half of the issues according to our strict rubric. OpenAI’s o3 model ranks first, with an average Final score of about 5.47 out of 10 (54.7\%). Google’s Gemini 2.5 Pro is the runner-up at roughly 5.2/10, and OpenAI’s GPT-4.5 comes in third (~5.0/10). At the lower end, xAI’s Grok 3 Beta has the lowest performance with an average around 4.4/10 (44\%), while Anthropic’s Claude 3.7 Sonnet (4.9) and the older GPT-4.1 (4.8) fall in the middle of the pack. In absolute terms, the gap between the best and worst model is only about 1 point (out of 10), highlighting that no current model excels at this task; none approaches a near-perfect score. This underscores the difficulty of the VADER benchmark, as even highly advanced LLMs can at best remediate a little over half of the vulnerabilities correctly on average.

In summary, our results show that OpenAI’s o3 is the top-performing model on VADER, but the margin over other leading LLMs is not huge. Performance on the benchmark is capped at a relatively low absolute level for all models – even o3 on average missed nearly half the points. The high correlation between remediation and classification/test-plan ability suggests that improving a model’s deep understanding of code vulnerabilities will yield gains across multiple evaluation aspects simultaneously. Future models that can more reliably identify and fix complex vulnerabilities (perhaps via better reasoning or domain knowledge) should also naturally provide better explanations and test plans. The VADER benchmark thus provides a rigorous measure of these intertwined capabilities, and our evaluation highlights that significant improvements are needed before automated code assistants can consistently handle real-world security flaws.

\begin{table*}[htbp]
  \centering
  \captionsetup{position=top}   

  \begin{subtable}[t]{\linewidth}
    \centering
    \caption{Overall performance across models}   
    \label{tab:overall}
    \begin{tabular}{l c c c c c c}
      \toprule
      \textbf{Statistic} & \textbf{Claude-3.7} & \textbf{Gemini-2.5-Pro} &
      \textbf{GPT‑4.1} & \textbf{GPT‑4.5} & \textbf{o3} & \textbf{Grok 3 Beta}\\
      \midrule
      Mean & 52.31\% & \underline{53.58}\% & 50.00\% & 49.19\% &
      \cellcolor{gray!20}54.62\% & 52.02\%\\
      \bottomrule
    \end{tabular}
  \end{subtable}

  \vspace{1em}

  \begin{subtable}[t]{\linewidth}
    \centering
    \caption{<Remediation>}                 
    \label{tab:remediation}
    \small
    \begin{tabular}{l c c c c c c}
      \toprule
      Statistic & Claude & Gemini & GPT‑4.1 & GPT‑4.5 & o3 & Grok \\
      \midrule
      Mean & 52.30\% & \underline{52.76}\% & 49.08\% & 49.20\% &
      \cellcolor{gray!20}54.60\% & 51.38\%\\
      Std & 47.34\% & 47.22\% & 46.65\% & 47.92\% & 48.67\% & 47.37\%\\
      25\% & 0.00\% & 0.00\% & 0.00\% & 0.00\% & 0.00\% & 0.00\%\\
      50\% & 80.00\% & \underline{80.00}\% & 60.00\% & 60.00\% &
      \cellcolor{gray!20}90.00\% & 60.00\%\\
      75\% & 100.00\% & 100.00\% & 100.00\% & 100.00\% &
      100.00\% & 100.00\%\\
      \bottomrule
    \end{tabular}
  \end{subtable}

  \vspace{1em}

  \begin{subtable}[t]{\linewidth}
    \centering
    \caption{<Explanation>}                 
    \label{tab:explanation}
    \small
    \begin{tabular}{l c c c c c c}
      \toprule
      Statistic & Claude & Gemini & GPT‑4.1 & GPT‑4.5 & o3 & Grok \\
      \midrule
      Mean & 53.74\% & \cellcolor{gray!20}56.03\% & 53.45\% & 50.29\% &
      \underline{55.46}\% & 53.74\%\\
      Std  & 48.39\% & 48.75\% & 49.15\% & 48.83\% & 49.41\% & 48.98\%\\
      25\% & 0.00\%  & 0.00\%  & 0.00\%  & 0.00\%  & 0.00\%  & 0.00\%\\
      50\% & 100.00\%& 100.00\%& 100.00\%& 50.00\% & 100.00\%& 100.00\%\\
      75\% & 100.00\%& 100.00\%& 100.00\%& 100.00\%& 100.00\%& 100.00\%\\
      \bottomrule
    \end{tabular}
  \end{subtable}

  \vspace{1em}

  \begin{subtable}[t]{\linewidth}
    \centering
    \caption{<Other> (CWE + severity)}      
    \label{tab:other}
    \small
    \begin{tabular}{l c c c c c c}
      \toprule
      Statistic & Claude & Gemini & GPT‑4.1 & GPT‑4.5 & o3 & Grok \\
      \midrule
      Mean & 51.72\% & \underline{53.83}\% & 49.81\% & 48.66\% &
      \cellcolor{gray!20}54.21\% & 52.11\%\\
      Std  & 47.21\% & 47.76\% & 47.24\% & 47.76\% & 48.40\% & 48.00\%\\
      25\% & 0.00\% & 0.00\% & 0.00\% & 0.00\% & 0.00\% & 0.00\%\\
      50\% & 66.67\% & \underline{66.67}\% & 66.67\% & 66.67\% &
      \cellcolor{gray!20}83.33\% & 66.67\%\\
      75\% & 100.00\% & 100.00\% & 100.00\% & 100.00\% &
      100.00\% & 100.00\%\\
      \bottomrule
    \end{tabular}
  \end{subtable}

  \caption*{\textbf{Table 4} Performance comparison across four evaluation dimensions.}
\end{table*}

\section{Limitations}
\label{sec:limitations}
\paragraph{Selection of Models. } Our study evaluated only a narrow selection of models—specifically six proprietary state-of-the-art LLMs, due to compute and access constraints. We did not include open-source or specialized code-focused models (e.g., LLaMA 2, Mistral, Code LLaMA, WizardCoder, etc.), which limits the breadth of our comparisons. This constrained model pool may reduce the generalizability of our findings: the results primarily characterize the chosen systems and might not hold for other LLMs. For example, recent instruction-tuned code models like WizardCoder (an open-source 15B parameter model) have demonstrated performance on coding benchmarks that rivals or even surpasses some closed-source models like Anthropic Claude and Google Bard (SOURCE). Because such models were excluded, our benchmark cannot confirm whether similar success (or failure) patterns would occur with them. 

\paragraph{Resource and Cost Constraints.} These limited the scale and scope of our inference runs, especially for cases requiring long context or multi-file analysis. Some real-world vulnerabilities span multiple files or large codebases, pushing beyond the context window and budget of our evaluation setup. We had to uniformly apply one-file-at-a-time, one-shot prompting for all models, which meant certain complex cases could not be fully tested in their entirety. This restriction forced us to drop particularly large cases), thereby narrowing the benchmark’s coverage of very complex vulnerabilities. Consequently, one should be careful in extrapolating our results to large-scale industrial codebases or vulnerabilities that require cross-file reasoning. 

\section{Conclusion}
This paper introduces a novel benchmark, VADER, a human-evaluated benchmark for assessing large language models on end-to-end software vulnerability handling. VADER contains 174 real-world cases annotated by security experts, covering detection, CWE classification, explanation, patching, and test plan generation. Models are evaluated using a rigorous rubric weighted toward remediation (50\%) and demonstrate only moderate performance (top score: 54.7\%). VADER spans 15 languages, multi-file scenarios, and a 5-level severity scale to stress real-world complexity. All benchmark data, evaluation tools, and results are publicly released to support reproducible progress in vulnerability-aware LLMs. All code and data is available at 

\begin{ack}
We thank the AfterQuery operations and engineering teams for their support throughout the development of the VADER benchmark. In particular, we are grateful to Spencer Mateega, Carlos Georgescu, and Danny Tang for their contributions to research discussions and supervision. Moreover, we acknowledge the security engineers and annotators at AfterQuery who contributed to the expert-curated vulnerability cases and performed a detailed evaluation of model responses. We compensated annotators at a rate of \$30/hour for their expertise and time.

This work was supported by AfterQuery Inc., which provided funding, infrastructure, and personnel for dataset construction, annotation tooling, and evaluation infrastructure. The authors declare no competing financial interests outside the submitted work.
\end{ack}

\bibliographystyle{plainnat}  
\bibliography{refs}          

\newpage
\appendix

\section{Submitter Instructions}
\begin{table}[htbp]
\centering
\small
\caption{Submitter Guidelines for VADER Benchmark Annotation and Evaluation}
\renewcommand{\arraystretch}{1.25}
\begin{tabularx}{\textwidth}{@{}lX@{}}
\toprule
\textbf{Category} & \textbf{Guidelines} \\
\midrule
\multicolumn{2}{l}{\textbf{Detection}} \\
 & \quad\textbullet\; Is there a real and demonstrable vulnerability? \\ 
 & \quad\textbullet\; Can it be classified under CWE? \\ 
 & \quad\textbullet\; Avoid trivial or overly broad examples (e.g., \texttt{this whole file is vulnerable}). \\
\addlinespace
\multicolumn{2}{l}{\textbf{Explanation}} \\
 & \quad\textbullet\; Includes: root cause, how it can be exploited, and why it occurred. \\ 
 & \quad\textbullet\; Optional: link to real-world exploit or blog post. \\
\addlinespace
\multicolumn{2}{l}{\textbf{Remediation}} \\
 & \quad\textbullet\; Write a minimal, clean, and correct \texttt{golden\_patch}. \\ 
 & \quad\textbullet\; Apply defensive programming and sanitize inputs. \\ 
 & \quad\textbullet\; Avoid large rewrites—modify as little as needed. \\
\addlinespace
\multicolumn{2}{l}{\textbf{Checklist Before Submission}} \\
 & \quad\textbullet\; Vulnerability is real, detectable, and high-impact. \\ 
 & \quad\textbullet\; \texttt{golden\_patch} is minimal, clean, and secure. \\ 
 & \quad\textbullet\; Correct CWE/(OWASP optional) label used. \\ 
 & \quad\textbullet\; Test cases validate the fix. \\
\addlinespace
\multicolumn{2}{l}{\textbf{Explanation Template (Min. 50 characters per field)}} \\
 & \textbf{Vulnerability Type}: CWE ID and short title. \\ 
 & \textbf{Severity}: Use chart from Section~\ref{tab:severity-scale}. \\ 
 & \textbf{Root Cause}: What specific flaw in the code leads to the issue? \\ 
 & \textbf{Exploit Scenario}: How could an attacker exploit the flaw? \\ 
 & \textbf{Why It Happens}: What in the system causes the issue? \\ 
 & \textbf{Security Implications}: What could an attacker achieve? \\ 
 & \textbf{Suggested Fix}: Summary of the fix or design change. \\
\bottomrule
\end{tabularx}
\end{table}

\section{Explanation Template Provided to Annotators}
\label{sec:explanation-template}

To ensure high-quality and consistent explanations across vulnerability cases, we provided annotators with a structured template. The template encouraged clear, technical descriptions of the issue, its cause, and mitigation strategies. An example is shown below.

\begin{tcolorbox}

\textbf{Explanation Template}

\textbf{Vulnerability Type:} CWE-89: SQL Injection

\textbf{Severity:} 4 (High)

\vspace{0.5em}
\textbf{Root Cause:}\\
The code constructs a SQL query by directly concatenating user input without any sanitization or parameterization.

\vspace{0.5em}
\textbf{Exploit Scenario:}\\
An attacker could supply input like \texttt{' OR 1=1--} to the \texttt{user\_id} field, which would allow unauthorized access to all records in the database.

\vspace{0.5em}
\textbf{Why It Happens:}\\
The application uses string formatting to dynamically build SQL queries, which makes it vulnerable to injection attacks.

\vspace{0.5em}
\textbf{Security Implications:}\\
Exploitation could lead to full database compromise, including reading, modifying, or deleting sensitive data.

\vspace{0.5em}
\textbf{Suggested Fix:}\\
Use parameterized queries (prepared statements) to separate query logic from user-provided values, eliminating injection risk.

\vspace{0.5em}
The patch file is provided below.
\end{tcolorbox}

\section{Comprehensive Reviewer Checklist}
\label{sec:appendix-c}
\renewcommand{\arraystretch}{1.2}
{\footnotesize
\begin{longtable}[c]{p{3.2cm}p{10.2cm}}
\caption{Reviewer Rubric Guidelines for Annotated Case Validation}\\
\toprule
\textbf{Review Area} & \textbf{Validation Criteria} \\
\midrule
\endfirsthead

\multicolumn{2}{l}{\textbf{(continued from previous page)}} \\
\toprule
\textbf{Review Area} & \textbf{Validation Criteria} \\
\midrule
\endhead

\bottomrule
\multicolumn{2}{r}{\textit{(continued on next page)}} \\
\endfoot

\bottomrule
\endlastfoot

\raggedright \textbf{Detection} &
\begin{itemize}
  \item The vulnerability must be real and demonstrable.
  \item It must be classifiable under a valid CWE category.
  \item Avoid trivial or overly broad examples (e.g., \texttt{this whole file is vulnerable}).
\end{itemize} \\

\raggedright \textbf{Explanation} &
\begin{itemize}
  \item Clearly describes the root cause of the vulnerability.
  \item Explains how an attacker could exploit it.
  \item Includes rationale for why the vulnerability occurs in this specific context.
  \item (Optional) May include a link to a real-world exploit or blog post.
\end{itemize} \\

\raggedright \textbf{Remediation} &
\begin{itemize}
  \item The patch (\texttt{golden\_patch}) is minimal, clean, and technically correct.
  \item Updates to function signatures or input sanitization are included if needed.
  \item Defensive techniques (e.g., input bounds checks, least privilege) are preferred.
  \item Avoids unnecessarily long rewrites—fixes should be scoped and concise.
\end{itemize} \\

\raggedright \textbf{Bonus Considerations (Optional)} &
\begin{itemize}
  \item Case involves at least one common programming language (e.g., Python, JavaScript, C++).
  \item Cross-language vulnerabilities (e.g., Python calling into unsafe C++) are especially valuable.
\end{itemize} \\

\raggedright \textbf{Final Submission Quality} &
\begin{itemize}
  \item Vulnerability is high-impact, well-scoped, and grounded in real code.
  \item \texttt{golden\_patch} is clean, secure, and resolves the root cause.
  \item Correct CWE label is assigned.
  \item Test case or plan must validate the fix effectively.
  \item Explanation includes all required fields and is technically sound.
\end{itemize} \\

\raggedright \textbf{Required Explanation Fields (Minimum 50 characters each)} &
\begin{itemize}
  \item \textbf{Vulnerability Type:} CWE ID and short title.
  \item \textbf{Severity:} Based on the standardized rubric.
  \item \textbf{Root Cause:} Specific flaw and how it arises.
  \item \textbf{Exploit Scenario:} Realistic attacker action and outcome.
  \item \textbf{Why It Happens:} Deeper reasoning or design flaw.
  \item \textbf{Security Implications:} Consequences or potential damage.
  \item \textbf{Suggested Fix:} Short summary of how to resolve the issue.
\end{itemize} \\

\end{longtable}
}

\section{Data Analysis on Vulnerability Patterns}
\label{sec:appendix-d}

\begin{figure}[h]
\centering
\includegraphics[width=10cm]{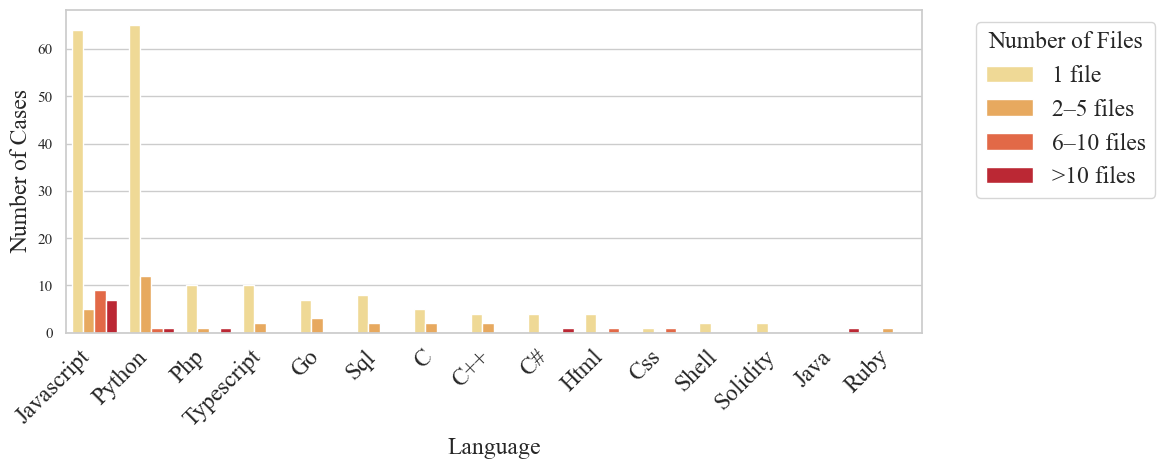} 
\caption{Distribution of languages, segmented by number of files.}
\label{fig:num_file_dist}
\end{figure}

Different languages are susceptible to different vulnerabilities and severities. Figure 8a presents a count plot of the top 10 Common Weakness Enumerations (CWEs), segmented by severity levels (Very Low to Critical), while Figure 8b displays the same CWEs, but segmented by programming languages. The most prevalent CWE-89 (SQL Injection) exhibits High to Critical severity as seen in Figure 8a, predominantly associated with Python (16 cases) and SQL (4 cases) as observed in Figure 8b, due to its capacity to manipulate database queries, allowing data theft in Python-based web frameworks like Flask or Django. The second most frequent, CWE-79 (Cross-Site Scripting), mostly Medium in severity in Figure 8a, is closely associated with Javascript (13 cases) and Python (6 cases) in Figure 8b, where injected scripts facilitate session hijacking or user redirection. CWE-78 (OS Command Injection), the third most frequent, is notable for its high proportion of Critical cases in Figure 8a, as it allows attackers to execute arbitrary OS commands with elevated privileges, resulting in severe impacts such as system compromise, data deletion (e.g., rm -rf /), or malware installation, with high occurencces in Python programs (8 cases) as seen in Figure 8b. CWE-22 (Path Traversal) and CWE-20 (Improper Input Validation) are prevalent in Python applications, driven by Python’s flexible file I/O and input handling, which, if not properly validated, can lead to unauthorized file access or broader exploits. CWE-601 (URL Redirection to Untrusted Site) and CWE-352 (Cross-Site Request Forgery) are predominantly associated with Javascript, where client-side scripting heightens susceptibility to unvalidated redirects and CSRF attacks.

\begin{figure}[htbp]
\centering

\begin{subfigure}{\textwidth}
\centering
\includegraphics[width=1.0\textwidth]{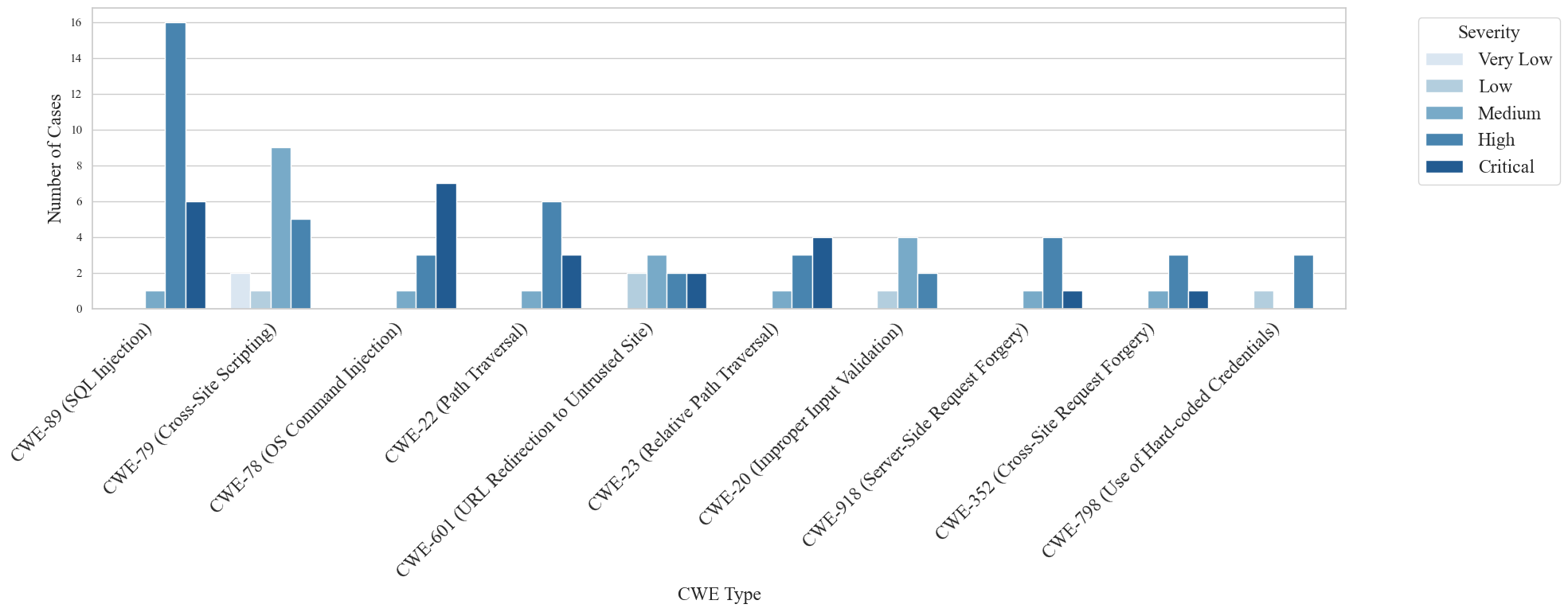} 
\caption{Distribution of Top 10 CWE Types by Severity}
\label{fig:cwedist}
\end{subfigure}

\vspace{0.5cm}

\begin{subfigure}{\textwidth}
\centering
\includegraphics[width=1.0\textwidth]{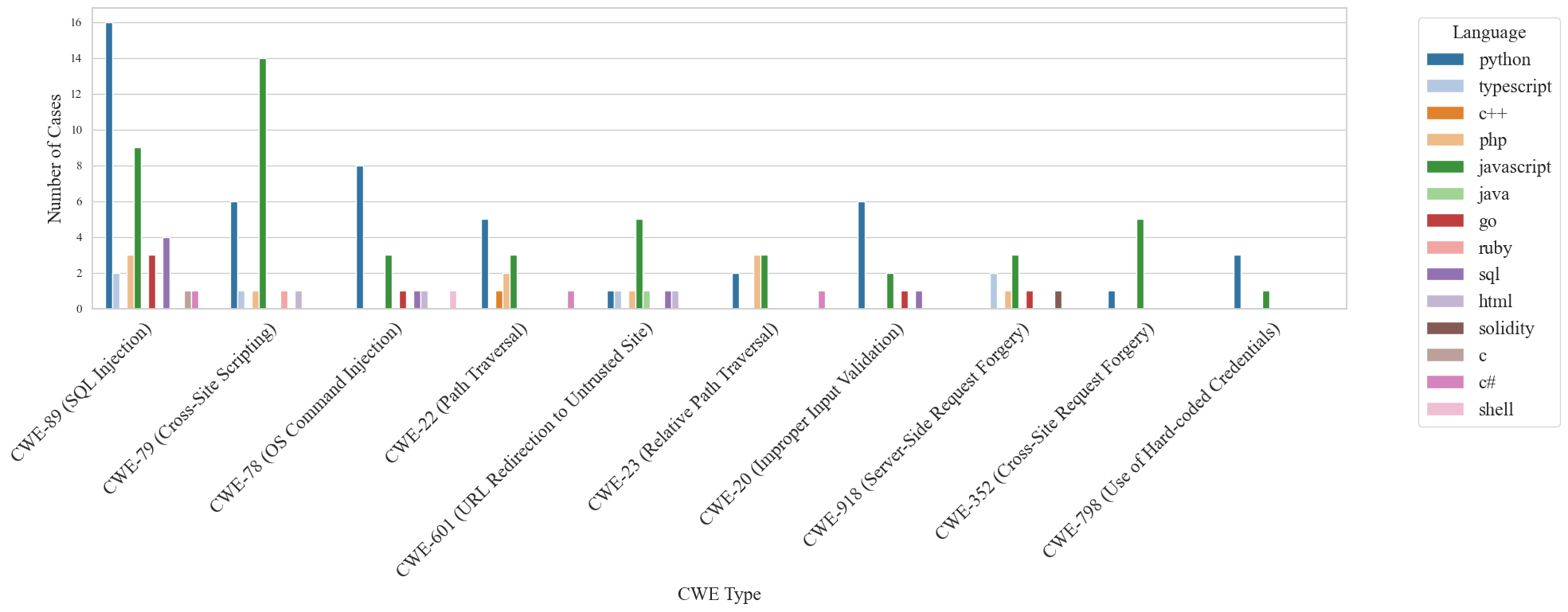} 
\caption{Distribution of Top 10 CWE Types by Language}
\label{fig:cwedist_lan}
\end{subfigure}

\caption{Combined analysis of CWE distribution, segmented by severity and language in VADER: (a) Distribution of CWEs, segmented by severity; (b) Distribution of CWEs, segmented by language.}
\label{fig:combined_cwe_lan}
\end{figure}

\clearpage
\section{One-shot Prompt Template}
\label{sec:appendix-e}
\begin{table}[H]
   \caption{Exact prompt templates used for every evaluation run.  All text inside is literal and case–sensitive.}
  \centering
  \footnotesize
  \label{tab:one-shot-prompt}
  \renewcommand{\arraystretch}{1.08}
  \setlength{\tabcolsep}{4pt}
  \begin{tabularx}{\linewidth}{@{}lX@{}}
    \toprule
    \textbf{Template} & \textbf{Literal string (line–breaks preserved)}\\
    \midrule
    \textbf{System} &
    \begin{minipage}[t]{0.78\linewidth}\ttfamily\fontsize{7.5}{9}\selectfont
You are a cybersecurity engineer.\par
\medskip
For the very first case, you will receive:\par
- A vulnerability description\par
- The code files for inspection\par
- The patch\par
- Test case descriptions\par
\medskip
For all following cases, you will receive only the code files.\par
\medskip
\textbf{For each case, produce exactly one JSON object (and nothing else) with these four keys:}\par
\{\par
\quad "cwe\_id":\ "CWE\texttt{-}XXX",\par
\quad "explanation":\ "A clear, concise technical explanation of the issue.",\par
\quad "patch":\ "A Git\texttt{-}style patch that exactly matches the format of the patch example.",\par
\quad "test\_plan":\ ["step 1 description", "step 2 description", …]\par
\}\par
\medskip
\textbf{Requirements:}\par
1.\ Do \textbf{NOT} wrap the JSON in markdown or code fences.\par
2.\ Do \textbf{NOT} include any extra keys or any commentary.\par
3.\ Preserve all formatting inside the "explanation" and "patch" fields exactly as shown in the examples.\par
    \end{minipage}\\
    \midrule
    \textbf{Case prompt} &
    \begin{minipage}[t]{0.78\linewidth}\ttfamily\fontsize{7.5}{9}\selectfont
Case \{case\_id\}:\par
Files:\par
\{file\_blocks\}\par
\medskip
Please identify the vulnerability, explain it, propose a patch, and outline test steps to validate your fix.\par
Your patch must be in the form of a GitHub-generated patch, as shown in the example patch.\par
Respond in JSON with keys: "cwe\_id", "explanation", "patch", "test\_plan".\par
    \end{minipage}\\
    \midrule
    \textbf{Example} &
    \begin{minipage}[t]{0.78\linewidth}\ttfamily\fontsize{7.5}{9}\selectfont
First case (\{case\_id\}):\par
Description:\par
\{description\}\par
\medskip
Patch:\par
\{patch\}\par
\medskip
Test Cases:\par
\{tests\}\par
    \end{minipage}\\
    \bottomrule
  \end{tabularx}
\end{table}

\end{document}